\documentclass[prd,twocolumn,showpacs,amsmath,amssymb,nofootinbib,superscriptaddress]{revtex4-1}

\usepackage{graphicx}% Include figure files
\usepackage{dcolumn}% Align table columns on decimal point
\usepackage{bm}% bold math
\newcommand{\Od}{{\cal O}}

\newcommand{\im}{\mbox{Im}\,}
\newcommand{\re}{\mbox{Re}\,}

\newcommand{\condtwo}{\langle \bar q q \rangle}

\newcommand{\condtwoT}{\langle \bar q q \rangle_T}
\newcommand{\conds}{\langle \bar s s \rangle}

\newcommand{\quarkcorT}{\langle {\cal T} (\bar q q)(x) (\bar q
q)(0)\rangle_{T}}

\newcommand{\be}{\begin{equation}}
\newcommand{\ee}{\end{equation}}
\newcommand{\ba}{\begin{eqnarray}}
\newcommand{\ea}{\end{eqnarray}}

\newcommand{\IR}{{\Bbb R}}

\newcommand{\gsim}{\raise.3ex\hbox{$>$\kern-.75em\lower1ex\hbox{$\sim$}}}
\newcommand{\lsim}{\raise.3ex\hbox{$<$\kern-.75em\lower1ex\hbox{$\sim$}}}

\begin{document}
%\baselineskip=20pt
% declarations for front matter

\title{Chiral Symmetry Restoration and Scalar-Pseudoscalar partners in QCD}
\author{A. G\'omez Nicola}
\email{gomez@fis.ucm.es}
\affiliation{Departamento de F\'{\i}sica
Te\'orica II. Univ. Complutense. 28040 Madrid. Spain.}
\author{J. Ruiz de Elvira}
\email{elvira@hiskp.uni-bonn.de}
\affiliation{Departamento de F\'{\i}sica Te\'orica II. Univ.
Complutense. 28040 Madrid. Spain.}
\affiliation{Helmholtz-Institut f\"ur Strahlen- und Kernphysik, Universit\"at Bonn, D-53115 Bonn, Germany}
\author{R. Torres Andr\'es}
\email{rtandres@fis.ucm.es}
\affiliation{Departamento de
F\'{\i}sica Te\'orica II. Univ. Complutense. 28040 Madrid. Spain.}

\begin{abstract}
We describe Scalar-Pseudoscalar partner degeneration at the QCD chiral transition  in terms of the dominant  low-energy physical states for the light quark sector. First, we obtain within  model-independent  one-loop Chiral Perturbation Theory (ChPT)  that the QCD pseudoscalar susceptibility is proportional to
 the quark condensate at low $T$. Next, we show that this chiral-restoring behaviour for $\chi_P$ is compatible with recent lattice  results for screening masses and gives rise to degeneration between the scalar and pseudoscalar  susceptibilities ($\chi_S$, $\chi_P$) around the transition point,  consistently with an  $O(4)$-like current restoration pattern. This scenario is clearly confirmed by lattice data when we compare $\chi_S(T)$ with the quark condensate, expected to scale as $\chi_P(T)$. Finally, we show that saturating $\chi_S$ with the  $\sigma/f_0(500)$ broad resonance observed in pion scattering and including its finite temperature dependence, allows to describe  the peak structure of $\chi_S(T)$ in lattice data and the associated critical temperature. This is carried out within a unitarized ChPT scheme which generates the resonant state dynamically and is also consistent with partner degeneration.
 \end{abstract}

\pacs{11.10.Wx, % Finite temperature field theory
11.30.Rd, % Chiral symmetries
12.39.Fe, % Chiral lagrangians
12.38.Gc. % Lattice QCD calculations
% 25.75.Nq. %Quark deconfinement, quark-gluon plasma production, and phase transitions
% 25.75.-q % Relativistic heavy-ion collisions
}

%\vspace{-.5cm}
%\rule{\textwidth}{.1mm}

\maketitle

\section{Introduction and Motivation}

Chiral symmetry breaking $SU_V(N_f)\times SU_A(N_f)\rightarrow SU_V(N_f)$ and its restoration, with $N_f$ light quark flavours, has been  a milestone in our present understanding of the Quantum Chromodynamics (QCD) phase diagram and  Hadronic Physics under extreme conditions of temperature $T$ and baryon density, as those produced  in Heavy-Ion and Nuclear Matter experimental facilities  such as RHIC, CERN (ALICE) and FAIR. Lattice simulations  support that deconfinement and chiral restoration take place very close to one another in the phase diagram. In the physical case $N_f=2+1$  ($0\neq m_u=m_d\equiv m_q\ll m_s$)  and for vanishing baryon chemical potential, they point towards a smooth crossover  transition at pseudocritical  temperature   $T_c\sim$ 145-165 MeV \cite{Aoki:2009sc,Bazavov:2011nk}, the results being fairly consistent with the $O(4)$  universality class \cite{Ejiri:2009ac}, which  would hold for two light flavours in the chiral limit $m_q=0$. The crossover nature of the transition means in particular that there is no unique way to identify the transition point, the most efficient one in lattice being the scalar susceptibility peak position, rather than the vanishing point for the quark condensate $\condtwoT$, the order parameter, which decreases asymptotically with  $T$ for $m_q\neq 0$.

The equivalence with the $O(4)\rightarrow O(3)$ breaking pattern for $N_f=2$ led to early proposals of  $\pi-\sigma$ meson degeneration (``chiral partners") at chiral restoration \cite{Hatsuda:1985eb} which in its simplest linear realization takes place through the $\sigma$-component of the $O(4)$ field $(\sigma,\pi^a)$ acquiring a thermal vacuum expectation value and mass both vanishing at the transition in the chiral limit. Degeneration in the  vector-axial vector sector ($\rho$ and $a_1$ states) as a signature of chiral restoration has also been thoroughly studied \cite{Rapp:1999ej}. Nowadays, we know that the $\sigma$ state is well established as a $\pi\pi$ scattering broad resonance for isospin and angular momentum $I=J=0$, known as $f_0(500)$ \cite{Beringer:1900zz}, which is then difficult to accommodate  as an asymptotically free state, like in the linear model. Precisely, one of our main conclusion here will be that this asymptotic description is not needed. In fact, in order to study chiral partner degeneration in the scalar-pseudoscalar sector, it is more appropriate to analyze the corresponding current correlation functions \cite{Shuryak:1996wx}  which can be derived from a chiral effective lagrangian without introducing explicitly a particle-like $\sigma$ degree of freedom.  The scalar and pseudoscalar susceptibilities in terms of  the corresponding QCD $SU(2)$ currents are given by:
\begin{widetext}
\begin{eqnarray}
\chi_S (T)&=&-\frac{\partial}{\partial m} \condtwoT=\int_E{d^4x \left[\quarkcorT-\condtwoT^2\right]}=\int_E{d^4x \left[\frac{\delta}{\delta s(x)}\frac{\delta}{\delta s(0)}Z[s,p]\bigg\vert_{s=m_q,p^a=0}\right]},\label{chisdef}\\
 \chi_P (T)\delta^{ab}&=&\int_E{d^4x \langle {\cal T} P^a (x) P^b (0)  \rangle_T}\equiv \delta^{ab} \int_E{d^4x K_P (x)}= \int_E{d^4x \left[\frac{\delta}{\delta p^a(x)}\frac{\delta}{\delta p^b(0)}Z[s,p]\bigg\vert_{s=m_q,p^a=0}\right]},\label{chipdef}
\end{eqnarray}
\end{widetext}
where $q=(u,d)$ is the quark field,  $P^a(x)=\bar q \gamma_5 \tau^a q (x)$ and $K_P(x)$ are respectively  the pseudoscalar current and its correlator, the Euclidean measure $\int_E{d^4x}=\int_0^\beta d\tau\int d^3\vec{x}$ with $\beta=1/T$ and $\langle \cdot \rangle_T$ denotes a thermal average.  For $\chi_P$,  parity invariance of the QCD vacuum ($\langle P^a \rangle_T=0$) and isospin symmetry have been used. In the above equation, $Z[s,p]$ is the QCD generating functional with scalar and pseudoscalar sources $(s,p^a)$ coupled to the massless lagrangian in the light sector as $-s(x) (\bar q   q) (x)+ip_a(x)P^a(x)$, so that $Z[m_q,0]$ is the QCD partition function.

Thus, should the scalar and pseudoscalar currents become degenerate at chiral restoration, $\chi_P(T)$ and $\chi_S(T)$ would meet at that point. Since $\chi_S$ is expected to increase, as a measure of the fluctuations of the order parameter, at least up to the transition point, it seems plausible that they meet near the transition. In an ideal $O(4)$ pattern, the matching should take place  near the maximum of $\chi_S$.

Since  $P^a$ has the quantum numbers of the pion field $\pi^a$, its correlators, like $K_P(x)$, are saturated by the pion state at low energies. Let us first review the prediction arising from the low-energy theorems of current algebra, equivalent to the leading order (LO)  in the low-energy expansion of chiral lagrangians. At that order,   one has $P^a\sim 2 B_0 F \pi^a$ (from PCAC theorem) with $B_0=M^2/2m_q$ and where $F$ and $M$  are the pion decay constant and mass respectively, so that $\chi_P\sim 4 B_0^2 F^2 G_\pi(p=0)+\dots$ from Eq.(\ref{chipdef}),  being $G_\pi(p)$ the pion propagator in momentum space $p\equiv(i\omega_n,\vec{p})$ and $\omega_n=2\pi n T$ the Matsubara frequency with integer $n$. Thus, the pseudoscalar correlator,  saturated  with the dominant pion state,  is just proportional to the pion propagator at this order. In addition, to LO the Euclidean propagator is just   the free one  $G_\pi^{LO}(p)=1/(-p^2+M^2)$ (interactions are suppressed at low energies)  with $p^2=(i\omega_n)^2-\vert\vec{p}\vert^2$, so that using also the Gell-Mann-Oakes-Renner (GOR) relation $M^2 F^2=-m_q\condtwo$, valid at this order, we would get $\chi_P\sim -\condtwo/m_q$, as a first indication of the relation between the pseudoscalar susceptibility and the quark condensate at the LO given by current algebra.

 The latter result can actually be obtained formally as a Ward Identity (WI) from the QCD lagrangian \cite{Broadhurst:1974ng}, in connection with the definition of the quark condensate for lattice Wilson fermions \cite{Bochicchio:1985xa}. However, both sides of the identity suffer from QCD renormalization ambiguities, so that this WI is formally well-defined only for  exact chiral symmetry \cite{Bochicchio:1985xa,Boucaud:2009kv}.   It is therefore interesting to study, and so we will do in the next section, how this identity is realized within Chiral Perturbation Theory (ChPT) \cite{Gasser:1983yg}, which describes the low-energy chiral symmetry broken phase of QCD in a model-independent framework where symmetry breaking is realized non-linearly and  pions are the only degrees of freedom in the lagrangian. The previous current-algebra results are actually just  the LO in the ChPT  expansion in powers of a generic low-energy scale $p$, denoting  pion momenta or temperature, relative, respectively, to $\Lambda_\chi\sim$ 1 GeV and $T_c$. These are nothing but  indicative natural  upper limits for the chiral expansion in terms of  scattering (typical resonance scale) and  thermodynamics (critical phenomena) respectively, although both are treated on the same foot in the chiral expansion. In particular, the LO prediction for $\chi_P$ is temperature independent, so it is not  obvious  that it can be simply extrapolated as, say, $\condtwo\rightarrow \condtwo_T$. Actually, all the quantities involved change with temperature due to pion loop corrections, namely $M_\pi (T)$, $F_\pi (T)$ and $\condtwo (T)$ \cite{Gasser:1986vb}.

Similarly, from Eq.(\ref{chisdef}), one can relate $\chi_S$ with the propagator of a "$\sigma$-like state" such that it couples linearly
to the external scalar source $s(x)$ in an explicit symmetry-breaking term  ${\cal L}_{SB}=2B_0 F s(x) \sigma (x)$.
Without further specification about its nature  and its coupling to other physical states such as pions,
one already gets $\chi_S\sim 4 B_0^2 F^2  G_\sigma(p=0)$, suggesting a growing behaviour inversely proportional to $M_\sigma^2$ as the
sigma state reduces its mass  to become degenerate with the pion.

We also recall that the problem of $\chi_S-\chi_P$ degeneration has been studied in nuclear matter at $T=0$  in \cite{Chanfray:2001tf}, to linear order in nuclear density. In that work, current algebra is assumed to hold  through PCAC in the operator representation and other low-energy theorems such as GOR,  which as discussed in the previous paragraphs, leads to the pseudoscalar correlator $K_P(x)$ being directly proportional to the pion propagator. The authors in \cite{Chanfray:2001tf} work within low-energy models at finite density for which this PCAC realization holds, so that by including the proper finite-density corrections to $G_\pi$, which carries out all the density dependence of $K_P$ through an in-medium mass, and to $\condtwo$, the relation $\chi_P\sim -\condtwo/m_q$ is found to hold in the nuclear medium. This result provides another supporting argument for the relation between the condensate and the pseudoscalar susceptibility and represents  an additional motivation for our present ChPT analysis, where we do not need to make any assumption about the validity of current algebra.

\section{Standard ChPT analysis of the pseudoscalar correlator and susceptibility}

The Next to Leading Order (NLO)  corrections to $\chi_P$ can be obtained systematically and in a model-independent way within ChPT, where one can also calculate  the scalar susceptibility $\chi_S$ to a given order only in terms of pion degrees of freedom.  The price to pay is that we expect to reproduce only the behaviour of $\chi_{S,P}(T)$ for low and moderate temperatures. However, since $\chi_P$ is dominated by pions, whose dynamics are well described through ChPT, we expect to obtain a reasonable qualitative description of its $T$ behaviour, whereas standard ChPT misses the peak structure of $\chi_S$ near the transition. We note in turn that
 the LO for $\chi_S$ vanishes, unlike that of $\chi_P$.  The ChPT NLO result for $\chi_S (T)$ can be found in \cite{Nicola:2011gq,GomezNicola:2012uc}.

 For $\chi_P (T)$ we consider the effective lagrangian ${\cal L}_2+{\cal L}_4+\dots$, where ${\cal L}_{2n}=\Od(p^{2n})$, including their dependence on the pseudoscalar  source $p^a$ as given in  \cite{Gasser:1983yg}. We follow similar steps as in \cite{GomezNicola:2012uc,GomezNicola:2010tb}, now for the pseudoscalar correlator $K_P(x)$.   The LO comes from ${\cal L}_2$ only and reproduces the current-algebra prediction. The NLO corrections to $K_P(x)$ are of the following types:

 \begin{itemize}

  \item The NLO corrections to the pion propagator $G_\pi^{NLO}(x)$, which come both from ${\cal L}_2$ one-loop tadpole-like contributions $G_\pi^{LO}(x=0)$ and from tree level ${\cal L}_4$ constant terms,

 \item Pion self-interactions  $\Od(\pi p^a \times \pi^3 p^b)$ in ${\cal L}_2$  contributing as $G_\pi^{LO} (x)G_\pi^{LO} (x=0)$,

    \item Crossed terms ${\cal L}_2=\Od(p^a \pi)  \times {\cal L}_4=\Od(p^b \pi)$ giving $G_\pi^{LO} (x)$ multiplied by a NLO contribution,

 \item ${\cal L}_4=\Od(p^a p^b)$ terms giving rise to a contact contribution proportional to $\delta^{(4)}(x)$.

\end{itemize}

The final result for the pseudoscalar correlator in momentum space for Euclidean four-momentum $p$ can be written as:

\begin{equation}
K_P (p)= a + 4B_0^2 F^2 G_\pi^{NLO} (p,T) + c(T) G_\pi^{LO} (p) +\Od(F^{-2}),
\label{Kp}
\end{equation}
where subleading terms are labeled by their $F^2$ dependence. The NLO propagator includes wave function and mass renormalization (at this order in ChPT there is no imaginary part for the self-energy):

\begin{equation}
G_\pi^{NLO}(p,T)=-\frac{Z_\pi (T)}{p^2-M_\pi^2(T)},
\label{nloprop}
\end{equation}
where the LO propagator corresponds to $Z_\pi=1, M_\pi=M$ and is temperature independent. $F$ and $M$ are the lagrangian pion mass and decay constant, related to the vacuum ($T=0$) physical values $M_\pi (0)\equiv M_\pi\simeq 140$ MeV, $F_\pi (0)\equiv F_\pi\simeq$ 93 MeV,
 by $\Od(F^{-2})$ corrections \cite{Gasser:1983yg}.

The constant $a$ in Eq.(\ref{Kp}) is temperature independent and is a finite combination of low-energy constants (LEC) of ${\cal L}_4$ \cite{Gasser:1983yg}. In the ChPT scheme, the divergent part of the ${\cal L}_4$  LEC cancels the loop divergences from ${\cal L}_2$ such that  pion observables are finite and independent of the low-energy renormalization scale. To NLO in ChPT all the pion loop contributions in Eq.(\ref{Kp}) are proportional to the tadpole-like contribution $G_\pi^{LO}(x=0,T)=G_\pi^{LO}(x=0,T=0)+g_1(M,T)$ with the thermal function:

\begin{equation}g_1(M,T)=\frac{T^2}{2\pi^2}\int_{M/T}^\infty dx  \frac{\sqrt{x^2-(M/T)^2}}{e^{x}-1},\label{g1}\end{equation}
which is an increasing function of $T$ for any mass.
Thus, the pion thermal mass in the NLO propagator is given at this order by $M^2_\pi (T)=M^2_\pi (0)\left[1+g_1(M,T)/2F^2\right]$ and is finite and scale-independent. The same holds for $F_\pi^2(T)=F_\pi^2(0)\left[1-2g_1(M,T)/F^2\right]$ and for $\condtwoT=\condtwo_0\left[1-3g_1(M,T)/2F^2\right]$ \cite{Gasser:1986vb}. Note  that  $M_\pi^{-2}(T)$ decreases with $T$ a factor of 3 slower than the condensate $\condtwoT$. In addition, to this order it holds $F_\pi^2(T)M_\pi^2(T)/\condtwoT=F_\pi^2(0)M_\pi^2(0)/\condtwo_0\neq -m_q$. That is, the GOR relation is broken at finite temperature to NLO by the same $T=0$ terms, given in \cite{Gasser:1983yg}. GOR holds to NLO only in the chiral limit, including temperature effects \cite{Toublan:1997rr}.

The constant $c(T)$ in Eq.(\ref{Kp}) includes both LEC contributions and loop functions and the same happens with the wave function renormalization constant $Z_\pi (T)$. Both are divergent, but the combination $4B_0^2 F^2 Z_\pi (T)+c(T)$ turns out to be finite and scale-independent. Note that if we replace $G_\pi^{LO}=G_\pi^{NLO}$ in the last term  in Eq.(\ref{Kp}), which is allowed at this order since $c(T)=\Od(F^0)$ is of NLO, that combination is precisely the one multiplying the NLO propagator, i.e, it is the $T$-dependent residue at the $M_\pi^2(T)$ pole (when the Euclidean propagator is analytically continued to the retarded one). That finite residue, being finite, has to be then a combination of the finite observables   involved. Actually, it happens to be:

\begin{eqnarray}
4B_0^2 F^2 Z_\pi (T)+c(T)=\frac{F_\pi^2(T)M_\pi^4(T)}{m_q^2}+\Od(F^{-2}).
\label{residue}
\end{eqnarray}

Thus, the expression in Eq.(\ref{residue}) represents the residue of the NLO $K_P$ correlator (\ref{Kp}) at the thermal pion pole. Note that by showing explicitly that the residue can be expressed as (\ref{residue}) we obtain that the thermal part of the pseudoscalar susceptibility $\chi_P(T)$ is the same as that in $-m_q\condtwo_T$, since $F_\pi^2(T)M_\pi^2(T)/m_q^2+\Od(F^{-2})=(\condtwoT/\condtwo_0)(F_\pi^2 M_\pi^2/m_q^2)+\Od(F^{-2})$, so that

\begin{eqnarray}\chi_P(T)&=&K_P(p=0,T)=K_P(p=0,T=0)\nonumber \\&+&\frac{F_\pi^2 M_\pi^2}{m_q^2}\frac{\condtwo_T-\condtwo_0}{\condtwo_0}+\Od(F^{-2}).
\label{Kpcond}
\end{eqnarray}

Now, since $\condtwo_T-\condtwo_0=\Od(F^0)$, at the NLO order we are working, we can replace in Eq.(\ref{Kpcond}) $F_\pi^2M_\pi^2=-m_q\condtwo_0+\Od(F^0)$ so that we get $\chi_P(T)-\chi_P(0)=-m_q(\condtwo_T-\condtwo_0)+\Od(F^{-2})$. Furthermore, the constant $a$ appearing in Eq.(\ref{Kp}) contains precisely the LEC combination that combines with that in the residue (\ref{residue}) to give the same scaling law, now including the $T=0$ part. Thus, our final result for the pseudoscalar susceptibility in ChPT to NLO (finite and scale-independent) is:

\begin{widetext}
\begin{eqnarray}
\chi_P^{ChPT}(T)&=&4B_0^2 \left[\frac{F^2}{M^2}+\frac{1}{32\pi^2}(4\bar h_1-\bar l_3)-\frac{3}{2M^2}g_1(M,T)\right]+\Od(F^{-2})
%\nonumber\\&=&
=-\frac{\condtwoT^{ChPT}}{m_q}+\Od(F^{-2}),
\label{chipchpt}
\end{eqnarray}
\end{widetext}
where the first term inside brackets is the LO current-algebra $\Od(F^2)$ and $\bar l_3$, $\bar h_1$ are renormalized scale-independent LEC \cite{Gasser:1983yg}.

Therefore, we have obtained the WI connecting $\condtwo$ and $\chi_P$ to NLO in model-independent ChPT, including finite-$T$ effects. Furthermore, the $m_q$ dependence   cancels in $\chi_P(T)/\chi_P(0)=\condtwoT/\condtwo_0$, where only meson parameters show up. To this order, $\chi_P(T)/\chi_P(0)=1-3g_1(M,T)/(2F^2)$ so that the LEC dependence also disappears. Recall that $\bar h_1$ comes from a contact term in ${\cal L}_4$ and is therefore another source of ambiguity in the NLO condensate \cite{Gasser:1983yg}.

Note also that, unlike the approach followed in \cite{Chanfray:2001tf}, we have arrived to the result (\ref{chipchpt}) without relying on the validity of current algebra, which actually holds only in ChPT to the lowest order. Actually, as explained, our result takes into account NLO corrections to $F_\pi (T), M_\pi(T), \condtwoT$ through GOR breaking terms, both at $T=0$ and $T\neq 0$, which turn out to be crucial to obtain the correct scaling law given in Eq.(\ref{chipchpt}).

\section{Lattice data analysis}

We can draw some important conclusions from the previous results.  First,  $\chi_P(T)$ scales like the order parameter $\condtwo$, instead of the much softer behaviour $1/M_\pi^2(T)$. This scaling suggests a chiral-restoring nature for the pseudoscalar susceptibility, although we cannot draw any definitive conclusion about chiral restoration just from our standard ChPT analysis, which makes sense only at low $T$. Its behaviour for higher $T$ approaching the transition should be considered merely as indicative extrapolations, pretty much in the same way as the ChPT prediction for the vanishing point of the quark condensate is just a qualitative indication that the restoring behaviour goes in the right direction. For this reason, in the following we will complement our standard ChPT calculation with a direct lattice data analysis, and later on with a unitarized study which, as we will see, incorporates the relevant degrees of freedom to achieve a more precise description near the transition point.

One can actually observe a clear signal of a critical chiral-restoring behaviour for $\chi_P$, consistent with our previous ChPT result, in the  lattice  analysis of Euclidean correlators,   which  determine their large-distance space-like screening mass $M^{sc}$ in different channels \cite{Cheng:2010fe}.   From  Eq.(\ref{chipdef}) we expect $\chi_P=K_P(p=0)\sim (M_P^{pole})^{-2}$ with $M_P^{pole}$ the pole mass associated to $K_P(p)$ in a general parametrization of the form $K_P^{-1}(\omega,\vec{p})=-\omega^2+A^2(T)\vert\vec{p}\vert^2+M_P^{pole}(T)^2$ with $A(T)=M_P^{pole}(T)/M^{sc}(T)$ \cite{Karsch:2003jg}. Here, $\omega=i\omega_n$ would correspond to the thermal Euclidean propagator and $\omega\in\IR+i\epsilon$ to the retarded Minkowski one setting the dispersion relation.
Assuming a soft temperature behaviour for $A(T)$, which is plausible below $T_c$ (for instance, $A=1$ for the NLO ChPT propagator in Eq.(\ref{nloprop})) we can then explain the sudden increase of  $M_P^{sc} (T)/M_P^{sc} (0)$ observed  in this channel \cite{Cheng:2010fe} since we expect that ratio to scale like  $\left[\chi_P(0)/\chi_P(T)\right]^{1/2}\sim \left[\condtwo_0/\condtwoT\right]^{1/2}$. We show in Fig.\ref{figlat} (left panel) these two quantities. The correlation between them  is notorious, given the uncertainties involved, and the mentioned increase is clearly observed. Data are taken from the same lattice group and under the same lattice conditions \cite{Cheng:2010fe,Bazavov:2009zn}. Note that in lattice works $(2m_q/m_s)\conds$  is subtracted from $\condtwo_T$ in order to avoid renormalization ambiguities. Estimating the $T=0$ condensates from NLO ChPT \footnote{For standard ChPT we use the same LEC values as in \cite{Nicola:2011gq,GomezNicola:2012uc}. Their influence is more important in $\chi_S$, due to the vanishing of the LO, than in $\chi_P$, $\condtwo$.}, this subtraction gives a 6\% correction and, from the lattice values,  it is about a 15\% correction near $T_c$. Apart from the screening versus pole mass and the strange condensate corrections, one should not forget about the typical lattice uncertainties, like resolution, choice of action, staggered taste breaking and large pion masses  \cite{Aoki:2009sc,Bazavov:2009zn}.

\begin{widetext}
\begin{center}
\begin{figure}
\centerline{\includegraphics[width=6cm]{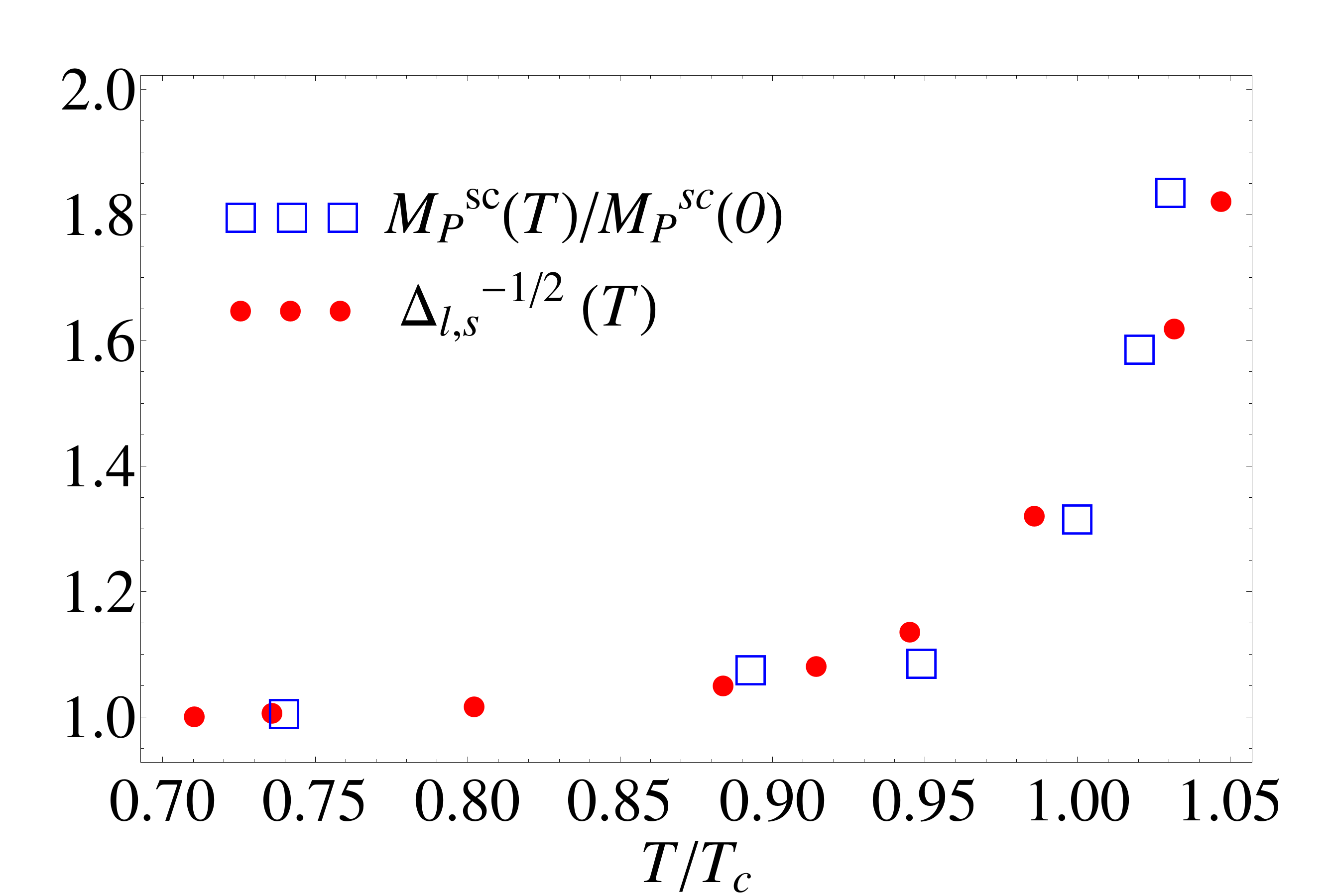}\includegraphics[width=6cm]{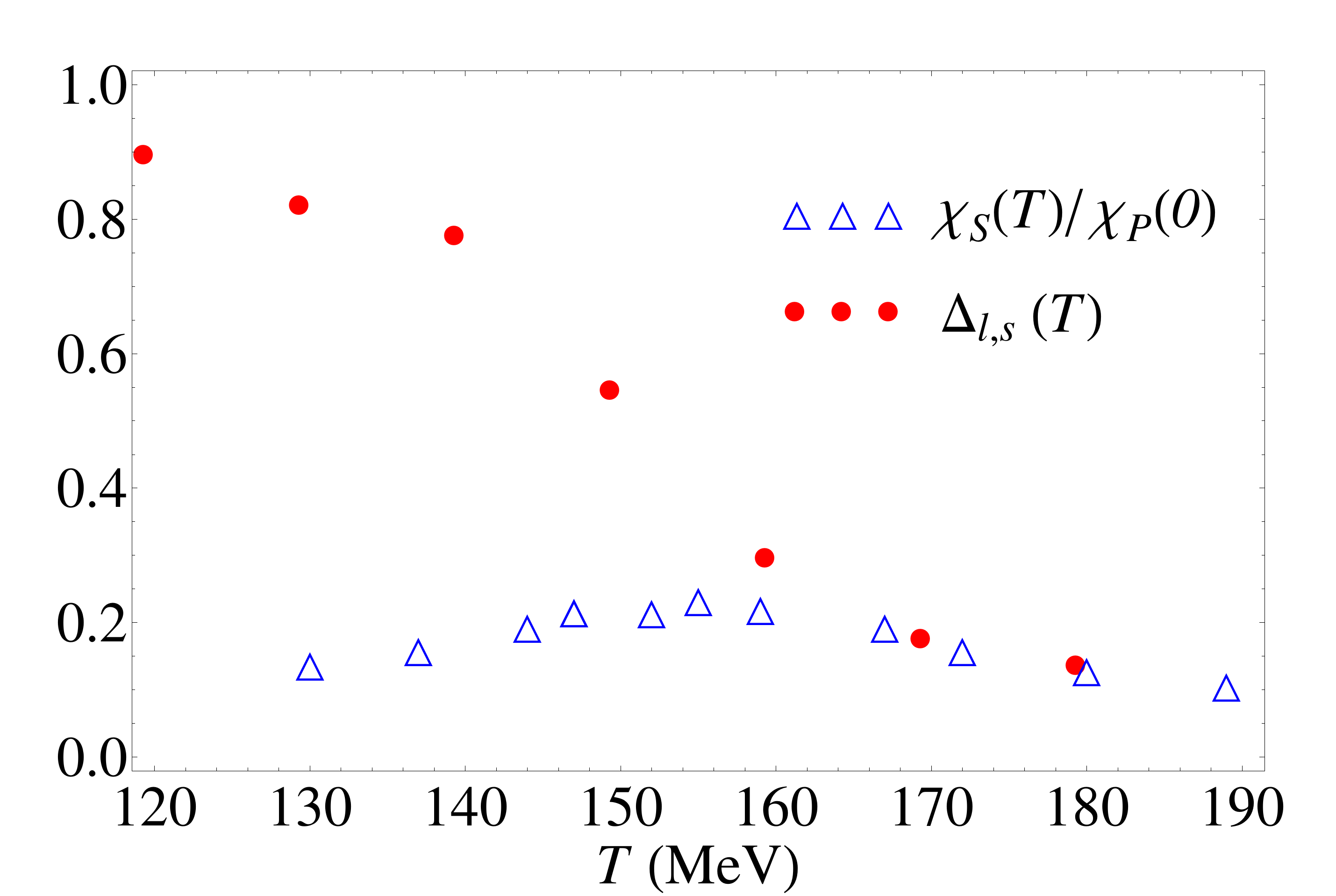}}
\caption{Left: Comparison between the pseudoscalar screening mass ratio and $\Delta_{l,s}^{-1/2}$, where $\Delta_{l,s}=r(T)/r(0)$ with $r=\condtwo-(2m_q/m_s)\conds$, for the  lattice data in
\cite{Cheng:2010fe} (masses) and \cite{Bazavov:2009zn} (condensate) with the same lattice action and resolution and $T_c\simeq 196$ MeV.
Right: Scalar susceptibility versus $\Delta_{l,s}\sim \chi_P(T)/\chi_P(0)$ from the data in \cite{Aoki:2009sc} for which $T_c\simeq 155$ MeV.}
\label{figlat}
\end{figure}
\end{center}
\end{widetext}

Another important conclusion of our analysis is that the decrease of $\chi_P$ and the increase of $\chi_S$ as they approach the critical point, lead to scalar-pseudoscalar susceptibility partner degeneration, which in an ideal $O(4)$ pattern should take place near the  $\chi_S$ peak. Once again, this behaviour is observed in lattice data. In Fig.\ref{figlat} (right panel) we plot the subtracted condensate, expected to scale as $\chi_P(T)/\chi_P(0)$ according to our previous ChPT and lattice analysis, versus $\chi_S(T)/\chi_P(0)$, both from the lattice analysis in \cite{Aoki:2009sc}. The $T=0$ values are taken from ChPT. The current degeneration is evident, not only at the critical point but also above it, where those two quantities remain very close to one another.

Recall that both the analysis of the correlation between screening masses and inverse rooted condensate and that of scalar versus pseudoscalar (condensate) susceptibilities, although elaborated from available lattice data, has not been presented before, to the best of our knowledge. As commented above, this analysis has been motivated by our  ChPT results in section II and  it gives strong support to the scalar-pseudoscalar degeneration pattern at the transition, as well as providing a  natural explanation for the behaviour of lattice masses in this channel.

\section{Unitarized ChPT and results}

\subsection{Extracting the $f_0 (500)$ thermal pole from  unitarized ChPT}

Since the scalar susceptibility is dominated by the $I=J=0$  lightest state, which does not show up in the ChPT expansion, let us consider its unitarized extension given by the Inverse Amplitude Method (IAM) \cite{Truong:1988zp} which generates dynamically in $SU(2)$ the  $f_0(500)$ and $\rho(770)$ resonances   and has been extended to finite temperature in \cite{GomezNicola:2002tn,Dobado:2002xf,FernandezFraile:2007fv}. Thus, before proceeding to derive the  unitarized susceptibility in section \ref{sec:unisus}, let us review briefly here, for the sake of completeness, some of the more relevant aspects of the thermal IAM, particularly in the scalar channel. We refer to  \cite{Dobado:2002xf,GomezNicola:2002an,GomezNicola:2002tn,FernandezFraile:2007fv} for a more detailed analysis.

\begin{figure}
\centerline{\includegraphics[width=8cm]{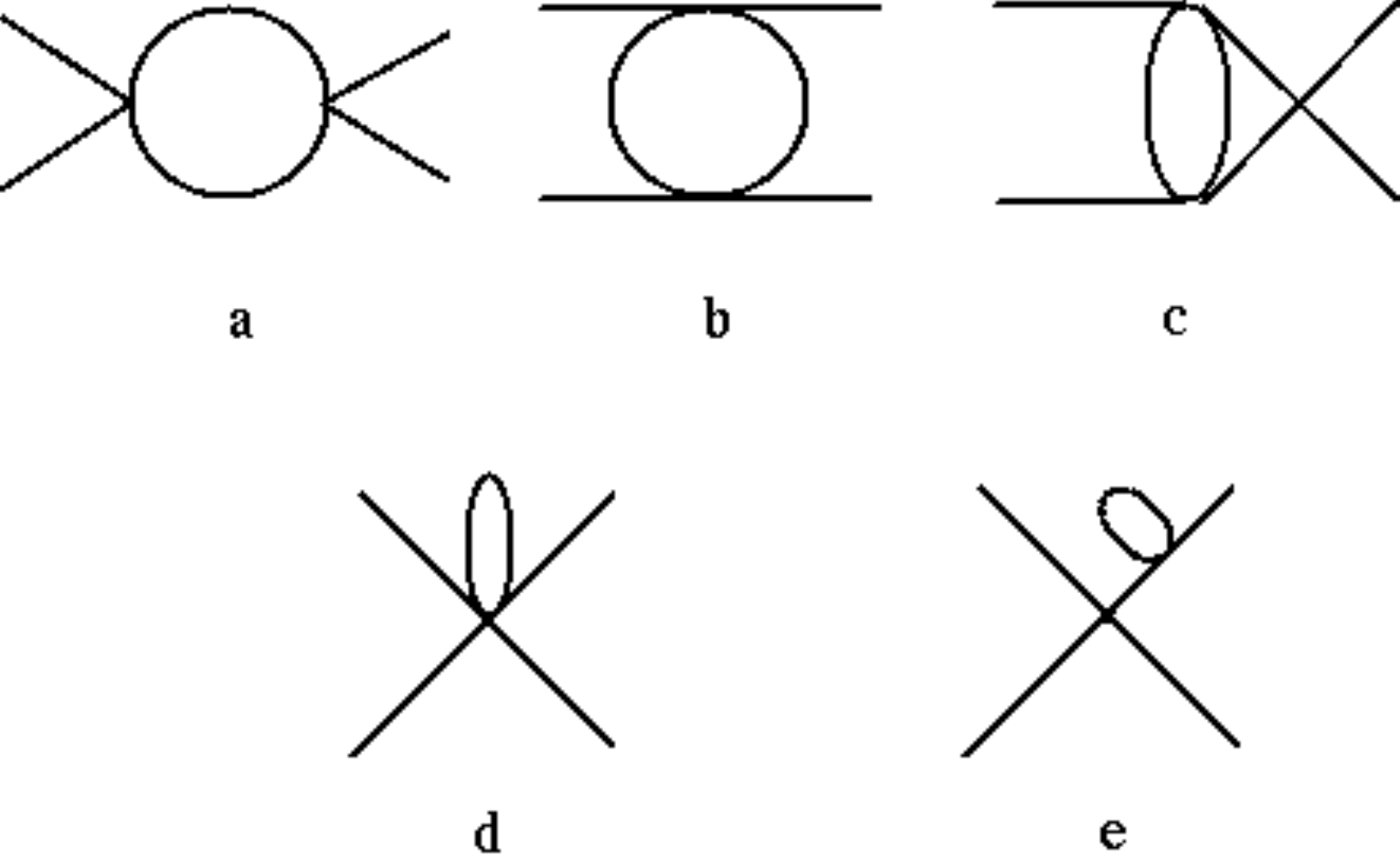}}
\caption{One-loop diagrams for  $T$-dependent pion scattering.}
\label{figfeyn}
\end{figure}

The IAM scattering amplitude is constructed by demanding unitarity and matching with the low-energy expansion, for which all the  ChPT scattering diagrams at finite temperature are included up to one-loop \cite{GomezNicola:2002tn}. The different types of those diagrams  are  represented in Figure \ref{figfeyn}.
The $T$-dependent corrections to the scattering amplitude comes from the internal loop Matsubara sums in the imaginary-time formalism of Thermal Field Theory. The external pion lines correspond to asymptotic $T=0$ states. The thermal amplitude is defined after the application of the $T=0$ LSZ reduction formula, which allows to deal just with thermal Green functions. After the Matsubara sums are evaluated, the external lines are analytically continued to real frequencies. The full result for the thermal amplitude to NLO in ChPT is given in \cite{GomezNicola:2002tn}.

The scattering amplitude can be projected into partial waves  $t_{IJ}(s)$ in the reference frame $\vec{p_1}=-\vec{p_2}$ where the incoming pions 1,2 are at rest with the thermal bath, so that $s=(E_1+E_2)^2$. The NLO partial waves have the generic form (we drop in the following the $IJ$ indices for brevity) $t(s;T)=t_2(s)+t_4(s;T)$ where $t_2(s)$ is the $\Od(p^2)$  tree-level $T$-independent scattering amplitude from ${\cal L}_2$ and  $t_4$
 is $\Od(p^4)$ including the tree level from ${\cal L}_4$ plus the one-loop from the diagrams in Fig.\ref{figfeyn}. Each partial wave satisfies $\im t_4 (s+i\epsilon;T)=\sigma_T(s) t_2(s)^2$ for $s>4M_\pi^2$ with $\sigma_T (s)=\sqrt{1-4M_\pi^2/s}\left[1+2n_B(\sqrt{s}/2;T)\right]$ and $n_B(x;T)=\left[\exp(x/T)-1\right]^{-1}$ the Bose-Einstein distribution.   This is the perturbative version of the unitarity relation for partial waves $\im t (s+i\epsilon;T)=\sigma_T (s) \vert t(s;T) \vert ^2$ and $\sigma_T$ is the two-pion phase space, which at finite $T$ receives the thermal enhancement proportional to $n_B$ which has a neat interpretation in terms of emission and absorption scattering processes allowed in the thermal bath \cite{GomezNicola:2002tn,GomezNicola:2002an}. Precisely imposing that the partial waves satisfy the above unitarity relation exactly while matching the ChPT series at low $s$ and low $T$,  leads to the thermal unitarized  IAM amplitude:
\begin{equation}
t^{IAM}(s;T)=\frac{t_2(s)^2}{t_2(s)-t_4(s;T)}
\label{iam}
\end{equation}

When the IAM amplitude is continued analytically to the $s$ complex plane \cite{Dobado:2002xf}, it presents poles in the second Riemann sheet $t^{II}(s;T)=t_2(s)^2/\left[t_2(s)-t_4^{II}(s;T)\right]$ with $t_4^{II}(s;T)=t_4(s;T)+2i\sigma_T t_2(s)^2$ so that $\im t^{II}(s-i\epsilon)=\im t^{IAM}(s+i\epsilon)$ for $s>4M_\pi^2$. Those poles correspond to the physical resonances, which in the case of pion scattering are the $f_0(500)$ ($I=J=0$) and $\rho(770)$ ($I=J=1$). The $T$-dependent poles can be extracted numerically by searching for zeros of $1/t^{II}(s;T)$ in the $s$ complex plane. We denote the pole position by $s_p (T)=\left[M_p(T)-i\Gamma_p(T)/2\right]^2$. The LEC for the IAM are chosen so that, within errors, they remain compatible with the standard ChPT ones and with the  $T=0$ pole values for the $\rho$ and $f_0 (500)$ listed in the PDG  \cite{Beringer:1900zz}.

Let us comment now on the thermal evolution of the resonance poles, whose main features for this work are represented in Figure \ref{figpoles}. Since $t_2(s)=a(s-s_0)$ with real $a$ and $s_0$, the thermal dependence  $s_p(T)$ is governed by that of $t_4^{II}$ at the pole.

\begin{widetext}
\begin{center}
\begin{figure}
\centerline{\includegraphics[width=6cm]{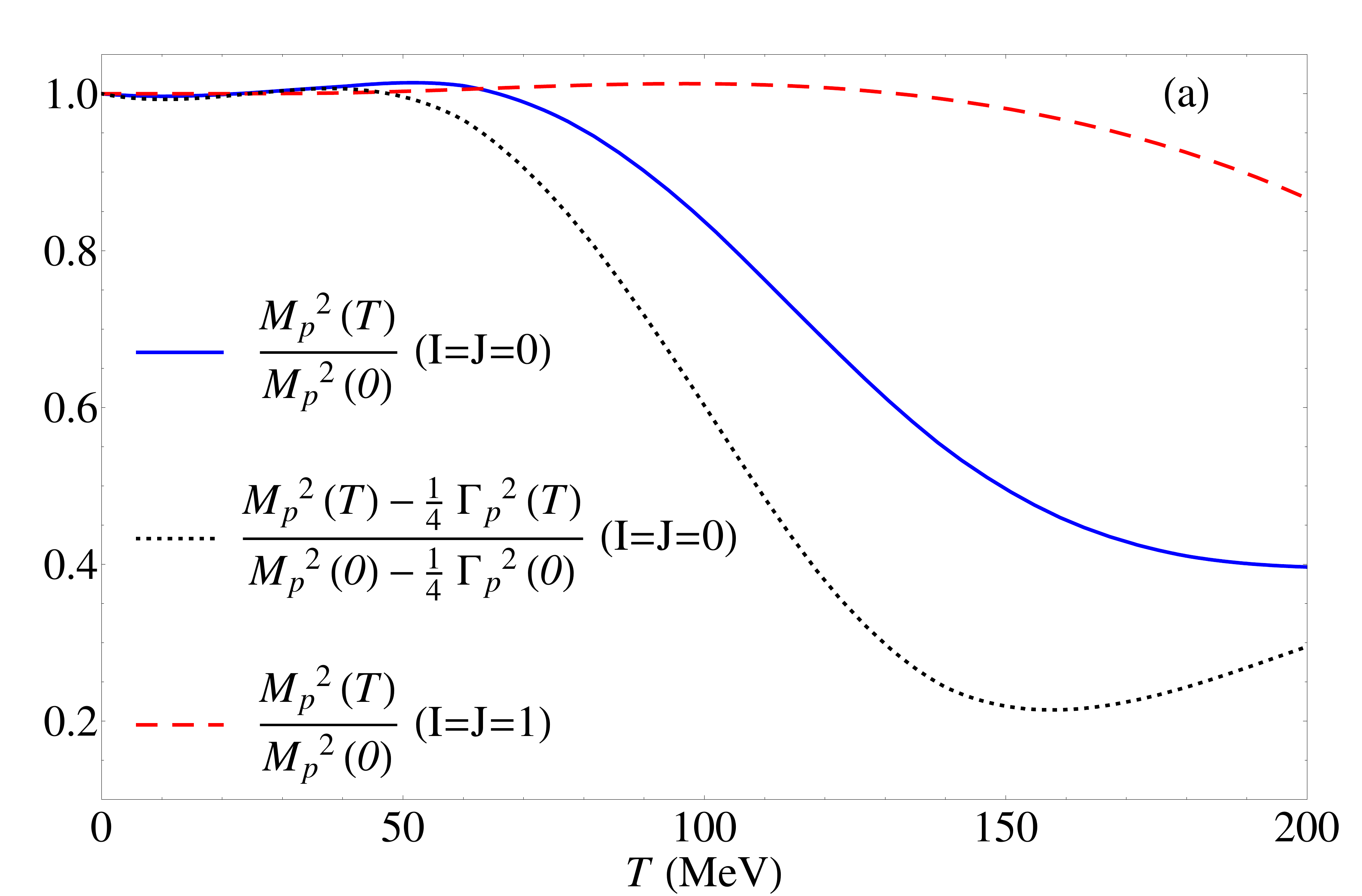}\includegraphics[width=6cm]{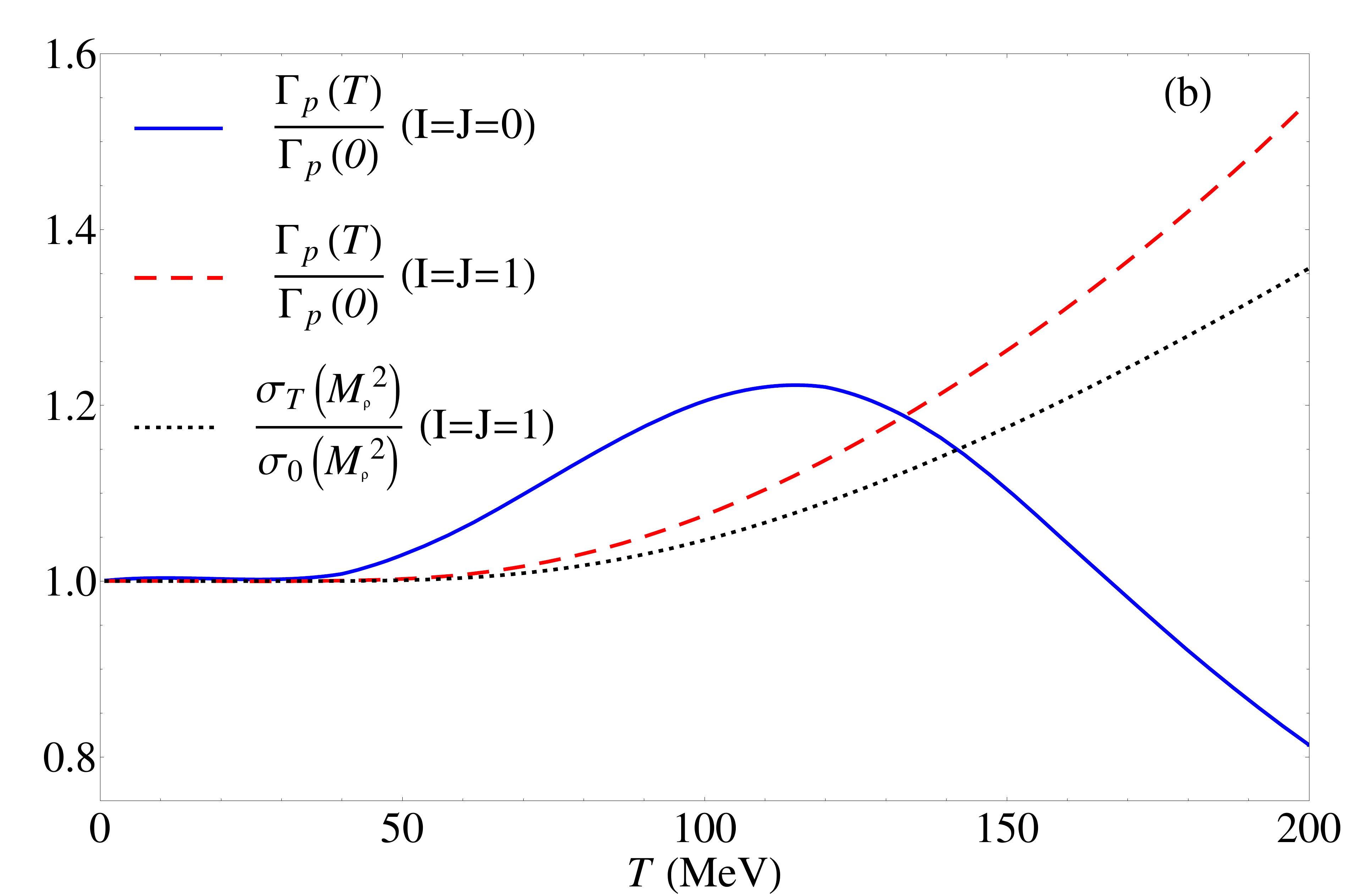}}
\centerline{\includegraphics[width=6cm]{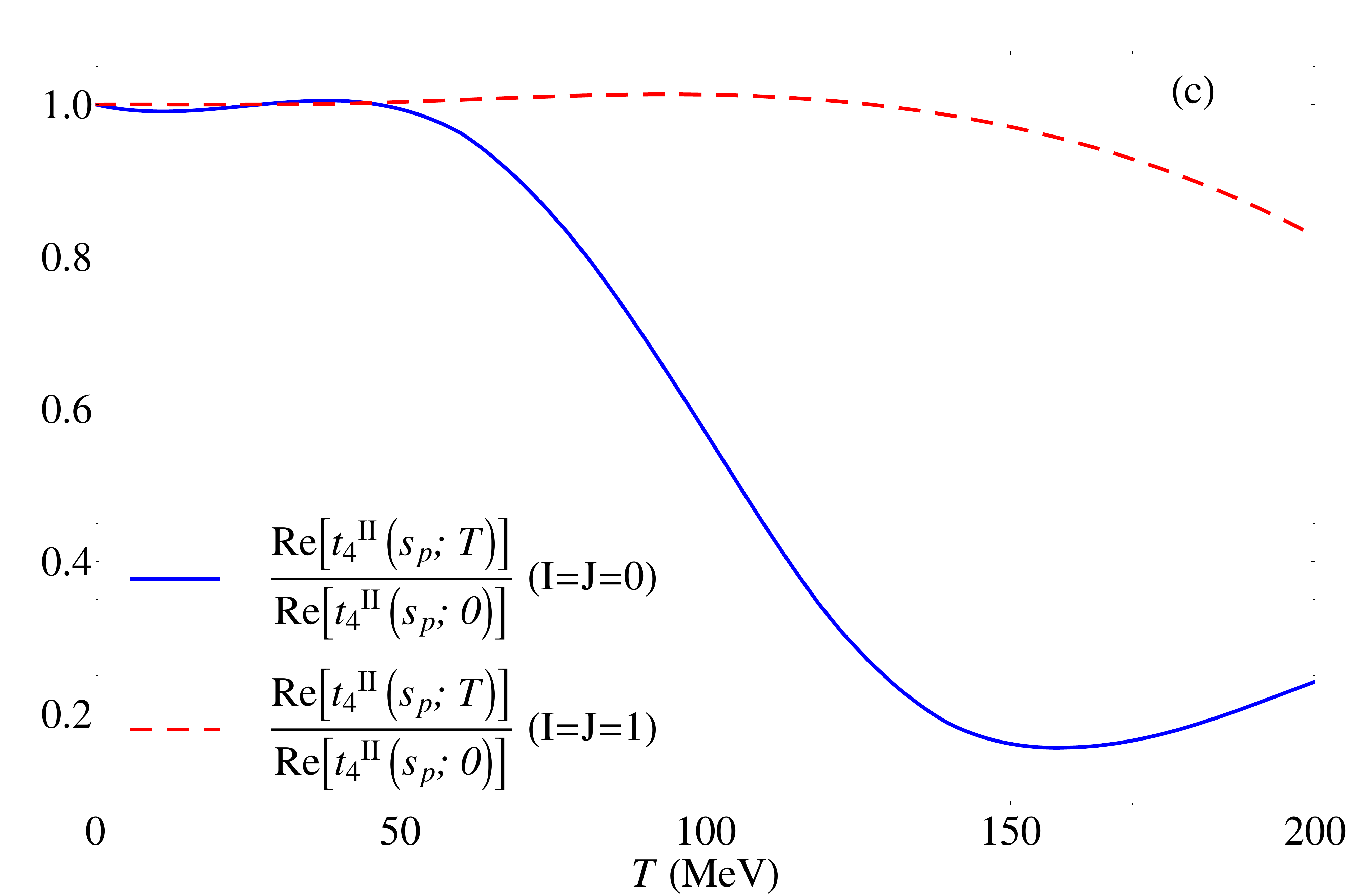}\includegraphics[width=6cm]{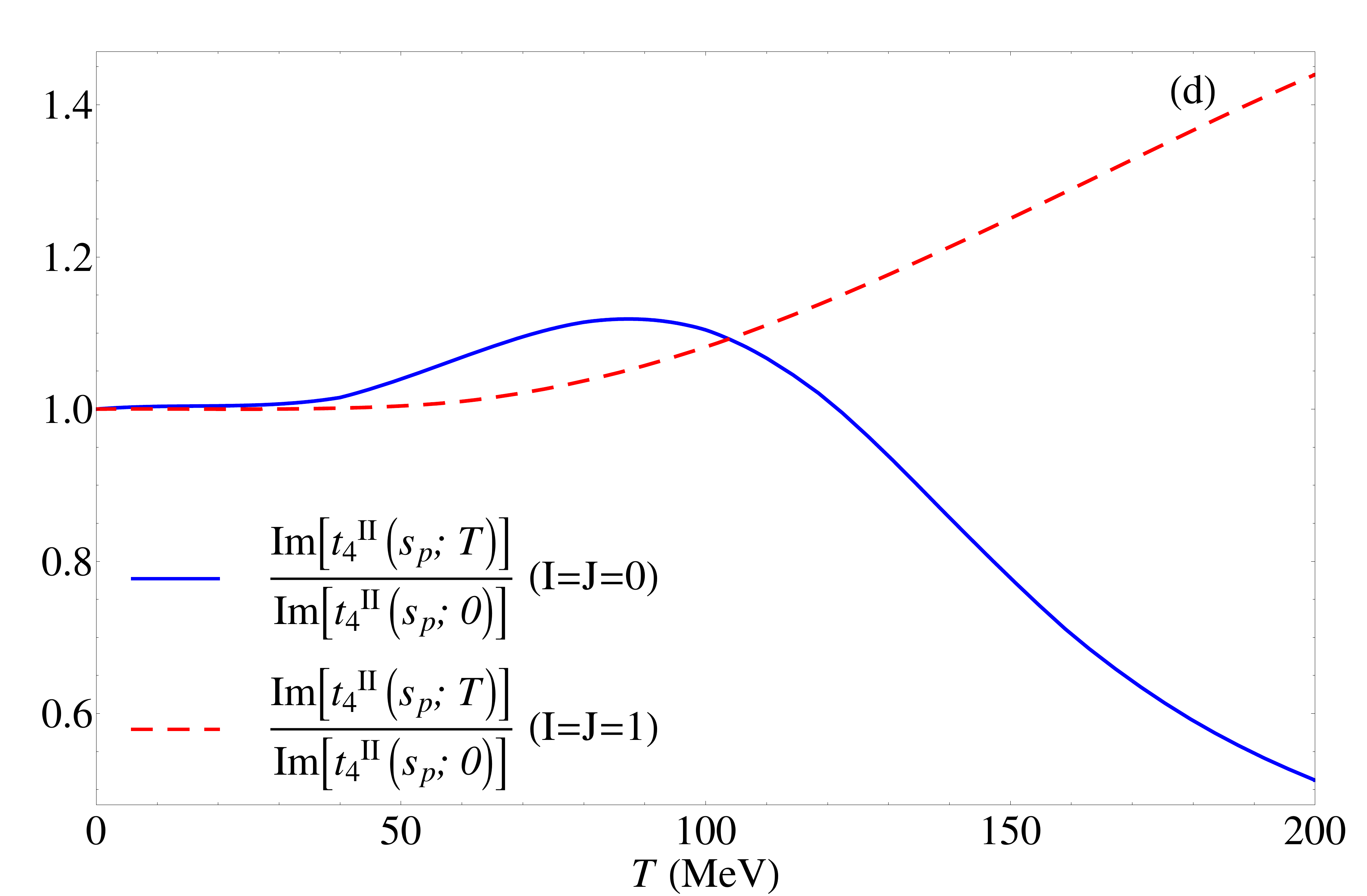}}
\caption{Thermal pole evolution (a)-(b) and contributions from the second sheet amplitude (c)-(d) for the scalar-isoscalar ($I=J=0$) and vector-isovector ($I=J=1$) channels.}
\label{figpoles}
\end{figure}
\end{center}
\end{widetext}

In the vector-isovector channel, $\Gamma_p<<M_p$ for all temperatures of interest here and therefore the $\rho$ can be considered a narrow Breit-Wigner (BW) resonance with $M_p$ and $\Gamma_p$ its mass and width respectively. Actually, $M_p^2(T)$ decreases very slightly with $T$ for the relevant  temperature range. Hence, the $T$-dependent contribution  of the real part of $t_4^{II}$ is almost negligible compared to its $T=0$ part, due to the large $\rho$ mass value. The latter gives roughly the $T=0$ rho mass, so the real part of the denominator of $t^{II}$ behaves dominantly as $s-M_p^2(0)$. In particular, the tadpole contributions of diagrams (d), (e) in Fig.\ref{figfeyn} are suppressed in this channel typically by $\Od(T^2/M_\rho^2)$. However, the thermal effect in $\Gamma_p(T)/\Gamma_p(0)$ is much more sizable, increasing with $T$. Its dominant contribution comes from the imaginary part of the amplitude. The imaginary part of the   $t^{II}$ denominator at the pole behaves like $M_p\Gamma_p(T)$ and, up to $T\simeq$ 100 MeV $\Gamma_p(T)/\Gamma_p(0)\sim\sigma_T/\sigma_0$, which for $s_p$ near the real axis comes essentially from diagram (a) in Fig.\ref{figfeyn} (which would give the only imaginary part for real $s>4M_\pi^2$ ) so the broadening can be explained just by thermal phase space increase up to that temperature. Note that, although this effect is  formally $\Od(e^{-M_\rho/T})$,  it activates below the transition because of the relative small value of $\Gamma_p(0)$. Above that, there is an additional increase of the effective $\rho\pi\pi$ coupling  with $T$ \cite{Dobado:2002xf}, to which tadpoles contribute, which explains a further increase of the width. This behaviour is represented in Figure \ref{figpoles}. Observe the softer behaviour of the mass as compared to the width in this channel, and the correlation with   the real and imaginary parts of the amplitude at the pole position.

In the scalar-isoscalar channel, the one we are interested in here, the behavior is remarkably different. We  rather talk of a broad resonance pole, since $M_p$ and $\Gamma_p$ are comparable, so that  the $f_0(500)$ pole is away from the real axis. As a consequence, all thermal contributions from  diagrams (a)-(e) in Fig.\ref{figfeyn} to $t_4^{II}$  become complex at the pole and  the real and imaginary parts of the pole equations do not have the simple form of a BW resonance. In addition, due to the lower value of $M_p^2(0)$  as compared to the $\rho$ case, the thermal dependence is much more stronger, both for the real and imaginary parts, and all contributions from those diagrams become equally relevant. In particular, the tadpoles in diagrams (d), (e) in Fig.\ref{figfeyn} now come into play. The numerical solution of the pole equations show that $M_p^2(T)$ in this channel decreases significatively, while $\Gamma_p (T)$ increases up to $T\simeq$ 120 MeV and decreases from that point onwards, as seen  in Fig.\ref{figpoles}. Note that this non-monotonic behaviour for $\Gamma_p$ cannot be explained now just in terms of phase space or vertex increasing. On the other hand, a possible interpretation of the decreasing $M_p^2$ is a chiral restoring behaviour. Actually, in Fig.\ref{figpoles} (a)  we also represent  $\re s_p (T)=M_p^2 (T)-\Gamma_p^2(T)/4\equiv M_S^2(T)$, which would correspond to the  self-energy real part of a scalar particle with energy squared $s$ and $\vec{p}=\vec{0}$, exchanged between the incoming and outgoing pions. This $\sigma$-like squared mass not only drops faster but  it develops a minimum at a certain temperature, which as we will see below corresponds to a maximum in the scalar susceptibility. In the $\rho$ channel, there is almost no numerical difference between $M_p^2$ and $\re s_p$. Thus, a qualitative explanation for the $\Gamma_p (T)$ change from a increasing  to a decreasing behaviour in the scalar channel would be the influence of the strong mass decreasing of the decaying state.

\subsection{Unitarized scalar susceptibility and quark condensate}
\label{sec:unisus}

In order to establish a connection between the scalar susceptibility and the scalar pole, we construct a unitarized susceptibility by
saturating the scalar propagator  with the $f_0(500)$ thermal state and assuming that its $p=0$ mass does not vary much with respect to the pole mass. Thus, we identify the pole of a scalar state exchanged in pion scattering with the thermal pole in the scalar channel discussed in the previous section. Therefore,  we have:
\begin{equation}
\chi_S^U(T)=\frac{\chi^{ChPT}_S (0) M_S^2(0)}{M_S^2(T)},
\label{susunit}
\end{equation}
where we have normalized to the $T=0$ ChPT value, since we are demanding that all our $T=0$ results match the model-independent ChPT predictions. This normalization compensates partly the difference between the $p=0$ and pole masses. Under this approximation, the  self-energy real part is  the squared scalar mass $M_S^2(T)=M_p^2(T)-\Gamma_p^2(T)/4$, as discussed in the  previous section, and the self-energy imaginary part vanishes at $p=0$.

The quark condensate cannot be extracted directly from the unitarized susceptibility. However,  we can obtain an approximate description by assuming that the relevant temperature and mass dependence, as far as the critical behaviour is concerned,  comes from  pion loop functions as $\delta\condtwo^U(T,M)=B_0 T^2 g(T/M)$ and $\delta\chi_S=B_0^2 h(T/M)$, with $\delta f(T)=f(T)-f(0)$. This $T/M$ dependence holds actually  to NLO ChPT, as in Eq.(\ref{g1}). Then, from Eq.(\ref{chisdef}), since $\delta\chi_S=-\partial \delta\condtwo/\partial m_q$, we  get
 \begin{equation} g(x)=g(x_0)+\int_{x_0}^x \frac{h(y)}{y^3} dy \qquad \mbox{for $x>x_0$},
 \end{equation}
 with $T_0=x_0 M\ll M$  a suitable low-$T$  scale below which we use directly NLO ChPT, which has a better analytic behaviour near $T=0$.  The $h$ function is obtained from the $T$ dependence of $\chi_S$ in Eq.(\ref{susunit}).

\subsection{Results}

Our theoretical results based on effective theories are plotted in Fig.\ref{figsusc}.  First, ChPT to NLO gives an increasing $\chi_S(T)$, intersecting   $\chi_P(T)$ at $T_d\simeq 0.9 T_c$, where $\condtwoT^{ChPT} (T_c)=\chi_P^{ChPT} (T_c)=0$. Once again, this result should be considered just as an extrapolation of the model-independent expressions for $\chi_S(T)$ and $\chi_P(T)$ beyond their low-$T$ applicability range. With this caution in mind, standard ChPT supports the idea of partner degeneration. Actually, near the chiral limit $M_\pi\ll T$, where  critical effects are meant to be enhanced, the degeneration point $T_d=T_c-3M_\pi/4\pi+\Od(M_\pi^2/T_c)$,  approaching the chiral restoration temperature  in that  limit.

\begin{widetext}
\begin{center}
\begin{figure}
\centerline{\includegraphics[width=10cm]{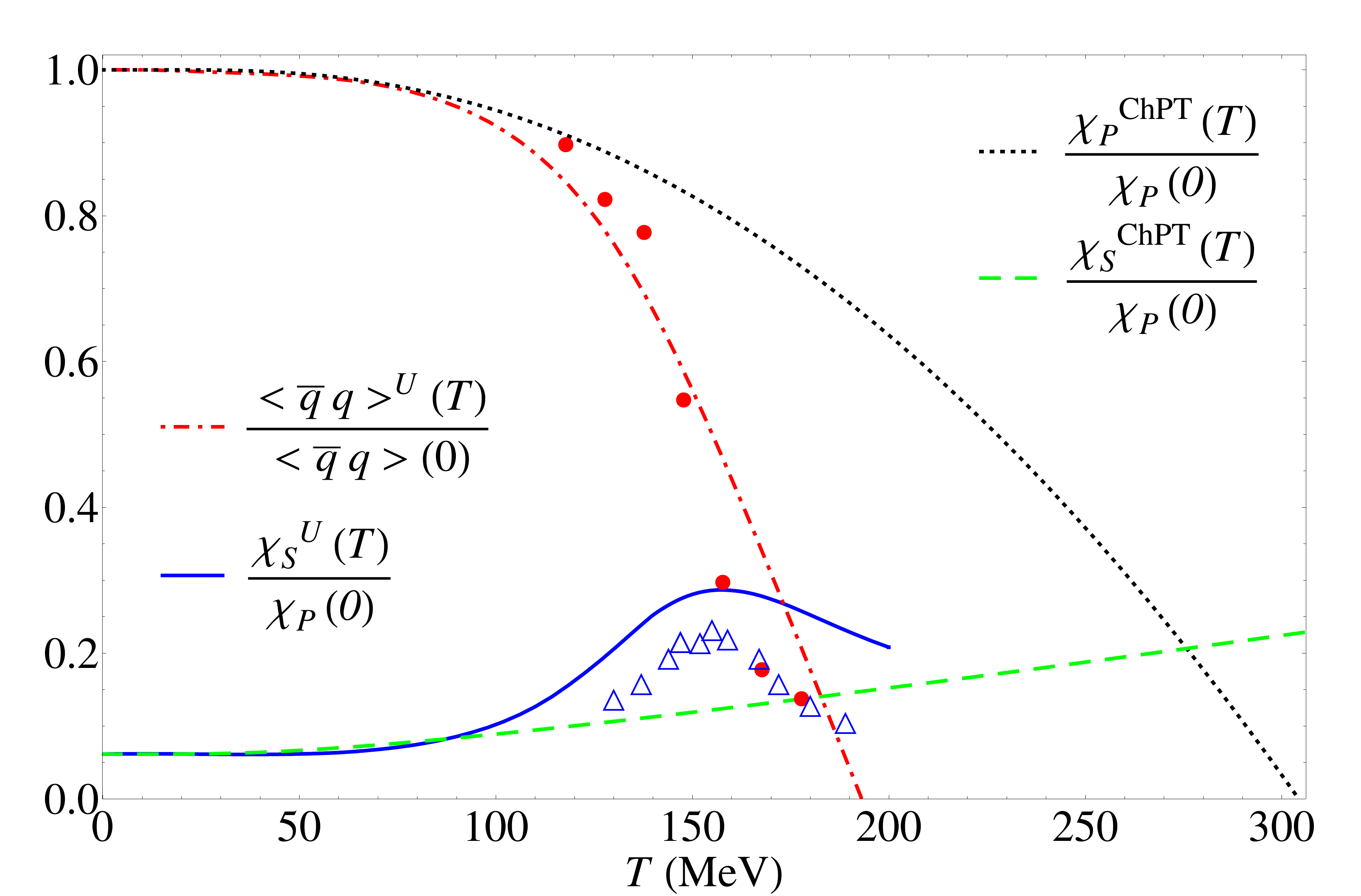}}
\caption{Scalar versus pseudoscalar susceptibilities  in ChPT and in our unitarized description. We show for comparison the lattice data of Fig.\ref{figlat} (right).}\label{figsusc}
\end{figure}
\end{center}
\end{widetext}

We also plot $\chi_S^U (T)$ in Fig.\ref{figsusc}.  The result agrees with standard ChPT  at  low $T$ and improves remarkably the behaviour near the transition. It actually develops a maximum at $T_c\simeq$ 157 MeV.  We show for comparison the lattice data of \cite{Aoki:2009sc}.  Furthermore, approaching the chiral limit by taking the $M_\pi=10$ MeV poles from \cite{FernandezFraile:2007fv} gives a vanishing $M_S(T)$ at $T_c\simeq$ 118 MeV and hence a divergent $\chi_S^U$ from Eq.(\ref{susunit}) at  $T_c$. Thus, we get, at least qualitatively, the  $T_c$ reduction and stronger $\chi_S$ growth near the chiral limit expected from theoretical \cite{Smilga:1995qf} and lattice \cite{Ejiri:2009ac,Bazavov:2011nk} analysis. The clear improvement of the unitarized approach with respect to the standard ChPT one for the scalar susceptibility is essentially due to the introduction of the thermal $f_0(500)$ state, whose importance in this case  is clearly seen from the dependence  $\chi_S\sim 1/M_S^2$, rather than to an enlargement of the applicability range in the unitarized approach. Actually, even within the unitarized description, we should be cautious when extrapolating it to near $T_c$ since we may be strictly beyond the effective theory range.

  The resulting $\condtwoT^U$ is  plotted in Fig.\ref{figsusc} with $T_0\simeq$ 12 MeV \footnote{We find very small numerical differences changing $T_0$ between 10-60 MeV.}. The critical behaviour is again nicely improved compared to the ChPT curves and is in better agreement with lattice data in that region. In addition, we obtain once more a scalar-pseudoscalar intersection near the $\chi_S$ peak and hence of chiral restoration. The corresponding $\condtwoT^U$ near the chiral limit ($M_\pi=10$ MeV) is much more abrupt, vanishing and meeting  $\chi_S$ at $T_c$ as expected.

Recall that in the above unitarized analysis, we are not performing a fit to lattice points. We just use the same LEC which generate the $T=0$ physical $f_0(500)$, $\rho$ states \cite{FernandezFraile:2007fv} and then provide our results for the susceptibility and condensate. The theoretical uncertainties in the LEC, as well as the lattice errors already discussed should be taken into account for a more precise comparison. Besides, our effective theory analysis is  not expected to reproduce the chiral restoration pattern above $T_c$.  In any case, apart from the important consistency obtained for chiral restoration properties, such as the $\chi_S$ peak and the $\chi_S/\chi_P$ matching, an  important point we want to stress is the crucial role of the $f_0(500)/\sigma$ state to describe the scalar susceptibility. Since $\chi_S\sim 1/M_S^2$, this observable is much more sensitive to this broad state, so that  a physically realistic description, including its thermal effects, turns out to be essential. This may not be the case for other thermodynamical observables, for which other approaches may provide a much better description. For instance,  the Hadron Resonance Gas (HRG) framework gives  very accurate results compared to lattice data \cite{Karsch:03,Huovinen:2009yb} and particle distributions \cite{Andronic:2008gu,Andronic:2012ut} below the transition, by including all known hadronic states as free particles in the partition function, often  including small interaction corrections. Within the HRG approach, hadron interactions are generically encoded in the resonant states. This framework works for most thermodynamic quantities, which are obtained, by construction, as monotonic functions of $T$, and generally increase with the mass of the states considered. For instance, the scalar susceptibility within that approach would increase with $T$, as it happens for other quantities such as the trace anomaly. Thus, the effect of a properly included broad $\sigma$ $T$-dependent state arising from pion scattering in order to describe chiral restoring properties such as the susceptibility peak, is once more highlighted. In fact, the $\sigma$
 state is just not included in many HRG works \cite{Huovinen:2009yb} or, at most, considered as a BW state \cite{Andronic:2008gu,Andronic:2012ut} with its $T=0$ mass and width, which, as we have commented above, does not provide an entirely adequate description. Finally, let us comment that apart from higher mass states, inclusion of higher order interactions may also be important for certain hadronic observables. For instance,  including  $\rho\pi$ interactions in the vector channel are essential to describe properly the dilepton spectra \cite{Rapp:1999ej}.

\section{Conclusions}

Summarizing, we have analyzed the scalar-pseudoscalar $O(4)$-like current degeneration pattern at chiral symmetry restoration,  from lattice simulations and  effective theory analysis. The pseudoscalar susceptibility scales as the quark condensate, which we have explicitly shown in ChPT to NLO at low temperature, and becomes degenerate with its ''chiral partner" scalar susceptibility  close to the scalar transition peak. The lattice data and the unitarized ChPT analysis support this picture and are well accounted for by the dominant physical states: pions and the $f_0(500)$ scalar resonance generated in pion scattering at finite temperature. In turn, we have provided a natural explanation for the sudden growth of lattice masses observed in the pseudoscalar channel. Although we have restricted here, for simplicity, to the $N_f=2$ case,  the analysis can be extended to  $N_f=3$, where the role of other scalar states such as the $a_0(980)$  can also be studied \cite{inprep}.

\section*{Acknowledgments}
Useful comments from F. Karsch, S. Mukherjee and D. Cabrera are acknowledged. Work partially supported by the EU FP7 HadronPhysics3 project,  the Spanish project  FPA2011-27853-C02-02 and FPI Programme (BES-2009-013672, R.T.A), and by the German DFG (SFB/TR 16, J.R.E.).

\end{document}